\documentclass{article}
\usepackage[T1]{fontenc}
\usepackage[preprint]{colm2026_conference}

\usepackage{microtype}
\usepackage{hyperref}
\usepackage{url}
\usepackage{booktabs}
\usepackage{graphicx}
\usepackage{amsmath,amssymb}
\usepackage{float}
\usepackage{xcolor}
\definecolor{citeblue}{rgb}{0.16, 0.42, 0.73}
\hypersetup{colorlinks=true, citecolor=citeblue, linkcolor=citeblue, urlcolor=citeblue}
\usepackage[capitalize,noabbrev]{cleveref}

\newcommand{\LambdaMed}{0.3}
\newcommand{\aHatMax}{1.97}
\newcommand{\aHatMed}{0.69}
\newcommand{\aHatMin}{0.13}
\newcommand{\aStarMed}{2.6}

\newcommand{\bestCfgView}{HM10, image normal, f/1.8}
\newcommand{\blurMedCm}{0.31}
\newcommand{\burstMaxMs}{500}
\newcommand{\burstMedianMs}{100}
\newcommand{\burstMinMs}{3}
\newcommand{\dMedAll}{3.6}
\newcommand{\dMedAllHi}{3.7}
\newcommand{\dMedAllLo}{3.4}

\newcommand{\dMedExpSeven}{3.4}

\newcommand{\fgwMax}{1.01}
\newcommand{\fgwMin}{0.14}
\newcommand{\fwhmPxCfgMax}{13}
\newcommand{\fwhmPxCfgMin}{8}
\newcommand{\ipBestMax}{1127}
\newcommand{\ipBestMin}{606}
\newcommand{\ipCandMin}{400}
\newcommand{\ipMax}{1127}
\newcommand{\ipMin}{402}

\newcommand{\lkAgreeMed}{+0.23}
\newcommand{\lkRatioMed}{0.13}
\newcommand{\lkShotRho}{+0.49}
\newcommand{\nArchiveShots}{15{,}969}
\newcommand{\nCampaigns}{4}
\newcommand{\nCfgPx}{6}
\newcommand{\nConfigs}{23}
\newcommand{\nConfigsOwn}{6}

\newcommand{\nExpSeven}{105}
\newcommand{\nExpTen}{142}
\newcommand{\nFramesProcessedM}{3.5}

\newcommand{\nOkBurst}{375}
\newcommand{\nOkCurrent}{421}
\newcommand{\nOkFlatTop}{404}

\newcommand{\nShotsBestCfgL}{37}

\newcommand{\nShotsCandidates}{368}

\newcommand{\nShotsFetched}{368}
\newcommand{\nShotsL}{183}
\newcommand{\nShotsOwnScale}{241}
\newcommand{\nShotsProcessed}{352}

\newcommand{\nShotsStats}{294}

\newcommand{\nStripShots}{468}
\newcommand{\nStripTenUs}{419}
\newcommand{\nTeAssumed}{234}
\newcommand{\nTeMeasured}{60}

\newcommand{\nTracksSel}{86{,}042}
\newcommand{\nWithinShot}{326}
\newcommand{\nXcorr}{175}
\newcommand{\outwardMed}{81}

\newcommand{\scaleCfgCampNine}{2.7}
\newcommand{\scaleCfgCampNineHi}{3.4}
\newcommand{\scaleCfgCampNineLo}{2.5}

\newcommand{\scaleCfgCampSix}{2.9}
\newcommand{\scaleCfgCampSixHi}{3.0}
\newcommand{\scaleCfgCampSixLo}{2.7}
\newcommand{\scaleCfgCampSixN}{70}
\newcommand{\scaleCfgCampSixRho}{-0.94}
\newcommand{\scaleCfgCampSixView}{HM10, image normal, f/1.8}
\newcommand{\scaleOwnMax}{5.2}
\newcommand{\scaleOwnMin}{2.7}
\newcommand{\scaleOwnRatio}{1.9}
\newcommand{\scaleOwnRhoStrong}{$-0.94$}
\newcommand{\scaleOwnRhoWeak}{$-0.58$}

\newcommand{\slBVFgw}{-0.10}
\newcommand{\slBVFgwHi}{+0.12}
\newcommand{\slBVFgwLo}{-0.26}
\newcommand{\slBVIp}{-0.29}
\newcommand{\slBVIpHi}{+0.66}
\newcommand{\slBVIpLo}{-1.02}

\newcommand{\slDFgw}{+0.06}
\newcommand{\slDFgwHi}{+0.21}
\newcommand{\slDFgwLo}{-0.08}
\newcommand{\slDIp}{+0.10}
\newcommand{\slDIpHi}{+0.40}
\newcommand{\slDIpLo}{-0.21}
\newcommand{\slDNbi}{+0.05}
\newcommand{\slDNbiHi}{+0.07}
\newcommand{\slDNbiLo}{+0.01}

\newcommand{\slVFgw}{-0.06}
\newcommand{\slVFgwHi}{+0.05}
\newcommand{\slVFgwLo}{-0.18}
\newcommand{\slVIp}{-0.05}
\newcommand{\slVIpHi}{+0.19}
\newcommand{\slVIpLo}{-0.29}
\newcommand{\slVNbi}{-0.01}
\newcommand{\slVNbiHi}{+0.01}
\newcommand{\slVNbiLo}{-0.03}
\newcommand{\slowFpsKhz}{50}
\newcommand{\teDefault}{30}
\newcommand{\vHatMed}{0.06}
\newcommand{\vMedAll}{424}
\newcommand{\vMedAllHi}{439}
\newcommand{\vMedAllLo}{406}

\newcommand{\vStarMed}{7345}

\newcommand{\vdRhoPooled}{+0.05}

\newcommand{\vdSlopeExpFour}{+0.07}
\newcommand{\vdSlopeExpFourHi}{+0.16}
\newcommand{\vdSlopeExpFourLo}{-0.04}
\newcommand{\vdSlopeExpSeven}{+0.05}
\newcommand{\vdSlopeExpSevenHi}{+0.11}
\newcommand{\vdSlopeExpSevenLo}{+0.01}
\newcommand{\vdSlopeExpTen}{+0.16}
\newcommand{\vdSlopeExpTenHi}{+0.27}
\newcommand{\vdSlopeExpTenLo}{+0.11}
\newcommand{\vdSlopeMed}{+0.11}
\newcommand{\vdSlopeMedHi}{+0.15}
\newcommand{\vdSlopeMedLo}{+0.08}
\newcommand{\vdSlopePooled}{+0.09}
\newcommand{\vdSlopePooledBlur}{+0.05}
\newcommand{\vdSlopePooledBlurHi}{+0.07}
\newcommand{\vdSlopePooledBlurLo}{+0.03}
\newcommand{\vdSlopePooledHi}{+0.11}
\newcommand{\vdSlopePooledLo}{+0.06}
\newcommand{\vdSlopePosShare}{70}
\newcommand{\vxPxCfgMax}{1.40}
\newcommand{\vxPxCfgMin}{1.15}
\newcommand{\windowMedianMs}{186}

\newcommand{\withinScaleMaxLog}{4}
\newcommand{\withinScaleMed}{5.8}

\newcommand{\withinScaleQHi}{13.6}
\newcommand{\withinScaleQLo}{3.8}
\newcommand{\xcorrRatioHi}{0.68}
\newcommand{\xcorrRatioLo}{0.52}
\newcommand{\xcorrRatioMed}{0.63}
\newcommand{\xcorrShotRho}{+0.29}
\newcommand{\xcorrVMed}{257}

\title{Scaling of the Radial Motion and Size of Filamentary\\Structures in the Scrape-Off Layer Inferred from 100~kHz\\Visible Imaging on \nShotsStats{} MAST Discharges}

\author{Jessie Dong$^{1}$ \qquad Ryan Wu$^{1,2}$ \\
$^{1}$Stanford University \\
$^{2}$SLAC National Accelerator Laboratory \\
\texttt{\{jessiedong, rwu4\}@stanford.edu}}

\begin{document}
\maketitle
\lhead{}
\renewcommand{\footrulewidth}{1.5pt}

\begin{abstract}
Filaments are important carriers of cross-field transport in the scrape-off layer of a tokamak. Analytic models predict that their radial velocity varies with their size. We measured the radial velocity and size of a large population of filaments on MAST, using data recorded from a 100~kHz, $48 \times 256$ pixel camera in the outboard midplane in the FAIR-MAST archive. Computer vision techniques were used to automate the identification of filaments and tracking of their trajectories, which yielded \nTracksSel{} filament tracks over \nShotsStats{} discharges in \nCampaigns{} campaigns. The archive does not contain spatial calibration, so we have obtained a pixel scale by regression of the plasma limb column in the image on the EFIT outboard LCFS radius, with bootstrap intervals. The scale is \scaleOwnMin--\scaleOwnMax{} millimetres per pixel.

We find a median velocity of \vMedAll{}~m\,s$^{-1}$, and median radial width of \dMedAll{}~cm. The dependence of velocity on size is mild, with a log--log slope of \vdSlopePooledBlur{} (95\% interval \vdSlopePooledBlurLo{} to \vdSlopePooledBlurHi) when the effects of exposure blurring are corrected for. This is because filaments are not in either velocity regime; they span the range $\hat a = \aHatMin$--$\aHatMax$ in two-region units, which straddles the maximum velocity predicted by the two-region model, but the normalized velocity $\hat v$ is much smaller than the scale of the model, at $\hat v \simeq \vHatMed$.

\end{abstract}

\section{Introduction}
\label{sec:intro}

Filaments are field-aligned structures of increased density and temperature that nucleate near the last closed flux surface (LCFS) in the edge of a tokamak and then move radially outward through the scrape-off layer (SOL). Filaments are responsible for a large part of the cross-field transport of particles and heat \citep{krasheninnikov2001scrape, zweben2007edge, dippolito2011convective}. The radial velocity of filaments determines the distance they travel before being drained to the divertor targets and the load on the first wall. Analytic models predict that the velocity of a filament depends on its size as well as parameters which control how quickly the polarization charge in the filament can be dissipated.

If the charge drains through the target sheath, then the velocity decreases with size as $v \propto a^{-2}$ \citep{krasheninnikov2001scrape}. If the circuit is closed by the ion polarization current instead, the velocity increases with size as $v \propto a^{1/2}$ \citep{garcia2006fluctuations}. These two limits are combined in the two-region model of \citet{myra2006collisionality}, which predicts that the velocity of the filament reaches a maximum at a critical size $a^*$ and falls on either side. The critical size depends on the sound speed, connection length and collisionality.

It is natural to test the validity of these models by measuring the velocities and sizes of a large number of filaments at known plasma conditions. The main diagnostic for this purpose on the Mega Amp Spherical Tokamak (MAST) is a Photron SA1.1 fast camera, which views the outboard midplane in visible light. Most of the analysis to date has used wide-angle views of the outboard midplane, and filaments have been identified either by hand or by software developed to detect filaments in that view. The analysis has been done on a small number of discharges \citep{dudson2008experiments, benayed2009interelm, kirk2016filaments, farley2019identification}, and it has been necessary to fit a model of the camera to get the spatial calibration.

The FAIR-MAST archive \citep{jackson2024fairmast} is focused on making the entire MAST data set public. One recording mode that earlier work has not used at scale is a narrow strip, 48 rows by 256 columns, running through the outboard edge of the plasma, which is recorded at 100{,}000 frames per second.

This strip image is capable of resolving the radial motion of filaments on 10~$\mu$s time intervals. There are only 48 rows in the image, however, which means the image contains information only on radial motion and therefore the radial size of the structures, and no information on other dimensions. The archive does not contain any spatial calibration, so we have to obtain the pixel scale from the data.

We have developed a pipeline to get statistics on the filaments from the raw images, based on standard computer vision techniques: background subtraction using a running minimum model, robust thresholding, detection of connected components, frame-to-frame tracking using the Hungarian algorithm, least-squares fit of tracks and, as an independent estimate of velocity, Lucas--Kanade optical flow. We calibrate the pixel scale against the equilibrium reconstruction (EFIT) by fitting the column in the image where the plasma limb occurs to the radius of the outboard LCFS reported by EFIT, since the limb column moves with this radius from shot to shot. The fit is done at the campaign level with a robust regression, yielding a value for the scale in millimetres per pixel. Bootstrap intervals are obtained for each camera configuration. The calibration uses only the images themselves and the EFIT data. We apply the pipeline to all the usable strip recordings in the FAIR-MAST archive.

The pipeline produces estimates of the radial velocity and radial full width at half maximum for each of the \nTracksSel{} tracks identified in the \nShotsStats{} MAST discharges. The tracks display a median radial velocity of \vMedAll{}~m\,s$^{-1}$ and a median radial FWHM of \dMedAll{}~cm. Eighty-one percent of the tracks are found to be moving outward. The slope of the log--log velocity--size relation is \vdSlopePooled{} before correction for exposure blur and \vdSlopePooledBlur{} for corrected widths, indicating near size-independence of the velocity over the range of observed filaments. In units of the two-region model, the filaments have normalized size of about $\hat a \approx \aHatMin$--$\aHatMax$ and normalized velocity $\hat v \approx \vHatMed$, which is at least an order of magnitude smaller than the characteristic velocity scale of the model.

We use robust regression to fit the dependencies of the median velocity and width on shot-averaged quantities such as plasma current, Greenwald fraction, and beam power, providing confidence intervals on the fitted dependencies. Code and data generated from the analysis are provided with this paper.

\section{Background}
\label{sec:background}

\paragraph{Filament motion.}
A filament is a coherent field-aligned tube of plasma, with a radius $a$ of the order of a few ion gyroradii to a few centimetres \citep{krasheninnikov2001scrape} and with greater density and temperature than the surrounding scrape-off layer (SOL) plasma. Magnetic curvature and gradient drifts in the outboard SOL transverse to the filament cause ions and electrons to separate, producing a polarization electric field that drives a radial $E \times B$ drift that moves the entire filament radially outward \citep{krasheninnikov2001scrape, dippolito2011convective}. The drift velocity is determined by the potential that results from the charge separation, and is sensitive to how the polarization charge is removed through the dissipation pathways. This leads to two limiting cases. When charge drains through the field-aligned sheath to the divertor target, the filament drifts at a speed $v \propto c_s (L_\parallel/R)(\rho_s/a)^2$ \citep{krasheninnikov2001scrape}, where $c_s$ is the sound speed, $\rho_s$ is the ion sound gyroradius, $L_\parallel$ the connection length, and $R$ the major radius. When the circuit is completed through the ion polarization current within the filament \citep{garcia2006fluctuations}, the drift speed is instead $v \propto c_s (a/R)^{1/2}$.

\paragraph{The two-region model.}
\citet{myra2006collisionality} generalized these limits, coupling the midplane and divertor leg in a two-region model to find the characteristic scale for the filament radius, the characteristic velocity scale, and the dimensionless width and collisionality,
\begin{equation}
    a^* = \rho_s^{4/5} L_\parallel^{2/5} R^{-1/5}, \qquad
    v^* = c_s \left(\frac{2 a^*}{R}\right)^{1/2}, \qquad
    \Theta = \left(\frac{a}{a^*}\right)^{5/2}, \qquad
    \Lambda = \frac{\nu_{ei} L_\parallel}{\Omega_e \rho_s},
    \label{eq:tworegion}
\end{equation}
where $\nu_{ei}$ is the electron--ion collision frequency and $\Omega_e$ is the electron gyrofrequency \citep{dippolito2011convective}. Defining the dimensionless width $\hat a = a/a^*$ and velocity $\hat v = v/v^*$,
\begin{equation}
    \hat v =
    \begin{cases}
        \hat a^{1/2} & \text{connected ideal interchange } (\Lambda < 1,\ \Theta < 1), \\
        \hat a^{-2} & \text{sheath connected } (\Lambda < 1,\ \Theta > 1), \\
        \Lambda\, \hat a^{-2} & \text{resistive X-point } (1 < \Lambda < \Theta), \\
        \hat a^{1/2} & \text{resistive ballooning } (\Lambda > 1,\ \Lambda > \Theta).
    \end{cases}
    \label{eq:regimes}
\end{equation}
At low collisionality, this predicts an increasing velocity with size as $\hat a^{1/2}$ for small filaments, a peak in the velocity around $\hat a = 1$, and a decreasing velocity with size as $\hat a^{-2}$ for large filaments. These limits are derived from scaling arguments, and more realistic simulations show slightly smaller velocities by factors of near unity due to details of filament profile and parallel structure \citep{easy2014three, walkden2015characterization, paruta2018blob}. Measurements of the velocity--size relation on the TORPEX basic plasma device \citep{theiler2009blob} found a transition between these predicted regimes.

\paragraph{Camera measurements.}
Visible imaging is sensitive to the line-integrated $D_\alpha$ emission, which depends on local density, temperature, and neutral density. Filaments show up as intense, field-aligned streaks riding on the background emission, and their emission is a close approximation of the filament density perturbation. Imaging-derived velocities and widths agree with probe measurements within tens of percent \citep{zweben2007edge, benayed2009interelm}. In the MAST tangential viewing geometry, the brightest feature in the camera image is the plasma limb, which is the region where the line of sight is tangent to the flux surfaces near the last closed flux surface (LCFS); filaments cross this limb as they move outward. Because the camera integrates over each exposure, a filament moving at velocity $v$ travels a distance $v\,t_{\rm exp}$ during the exposure, smearing the image of the filament over this distance in the direction of its motion (Section~\ref{sec:analysis}).

\section{Data}
\label{sec:data}

The FAIR-MAST archive \citep{jackson2024fairmast} contains the MAST dataset of \nArchiveShots{} discharges, with each discharge stored in a separate Zarr store and accessible via HTTP. The database includes a metadata API for the discovery of available signals and shot parameters. The groups that we access from the archive are the midplane fast camera (\texttt{rbb}), the EFIT reconstruction (\texttt{efm}), the summary signals for line-averaged and Greenwald density (\texttt{esm}), the interferometer (\texttt{ane}), the core and edge Thomson scattering (\texttt{ayc}, \texttt{aye}), and the midplane $D_\alpha$ monitor (\texttt{xim}). In addition to these signals, we also access per-shot metadata including the plasma current, heating scheme, injected beam power and the interval of the flat top.

\subsection{Fast camera (\texttt{rbb})}
The \texttt{rbb} signal corresponds to the Photron SA1.1 camera (also known as the ``bullet camera B'') situated at the HM10 midplane port and aimed tangentially at the outboard edge \citep{kirk2016filaments, farley2019identification}. The recorded region is $48 \times 256$ pixels and recorded at 100{,}000 frames/s with 10 bits resolution. The offset of this region on the sensor is stored as part of the per-shot metadata. The exposure times vary between 1 and 10~$\mu$s, and the most common settings are 4, 7, and 10~$\mu$s. Also recorded are the lens, aperture, and filter settings and whether the camera was inverted. We do not rely on the orientation of the camera for our detection pipeline (Section~\ref{sec:method}), but all camera configurations are catalogued in Appendix~\ref{app:views}, along with the details of their calibrations.

\subsection{Discharge selection}
\nStripShots{} discharges from FAIR-MAST have the fast camera recording a $48 \times 256$ pixel region. These are in the form of bursts, high-speed recordings made in a window within a slower recording of the whole discharge at 1--2~kHz. The median duration of these bursts is \burstMedianMs{}~ms, with extremes of \burstMinMs{} to \burstMaxMs{}~ms. Of these, \nStripTenUs{} have frame intervals of 10~$\mu$s. We select discharges for further processing with the following criteria: (i) the burst of interest overlaps the plasma flat top, extended by 20~ms either side, by at least 2{,}000 fast frames. We find \nOkBurst{} of these discharges. (ii) the flat top must be defined and at least 50~ms long. We find \nOkFlatTop. (iii) a peak plasma current of at least \ipCandMin{}~kA. We find \nOkCurrent{} discharges, and \nShotsCandidates{} discharges satisfy all three conditions.

For the \nShotsFetched{} selected discharges, we download and process the frames that are in the burst and between the extended flat top. The median length of this processing window is \windowMedianMs{}~ms, leading to a total of \nFramesProcessedM{} million frames processed. Sixteen of the selected discharges have a slower burst rate of \slowFpsKhz{}~kHz; these are included in the processing, but are excluded from the statistics to avoid changing time resolution. The set of processed discharges is shown broken down by campaign in Table~\ref{tab:campaigns}, with the number of discharges, exposure settings, pixel scales, and median velocity and width for each.

\begin{table}[t]
\centering
\scriptsize
\setlength{\tabcolsep}{4pt}\begin{tabular}{llrrllllr}
\toprule
Campaign & Years & Shots & In stats & Exp. ($\mu$s) & Pooled scale (mm/px) & Own scales & $\tilde v_r$ (m/s) & $\tilde\delta_r$ (cm) \\
\midrule
M6 & 2006--2007 & 257 & 220 & 1, 5, 7, 10 & 3.7 (3.3--4.1) & 3 (2.9--3.7) & 419 & 3.6 \\
M7 & 2008--2009 & 15 & 13 & 4, 10 & 3.3 (3.0--3.8) & 1 (3.3) & 451 & 3.6 \\
M8 & 2011 & 32 & 25 & 2, 4 & 5.2 (3.6--11.5) & 1 (5.2) & 627 & 5.2 \\
M9 & 2013 & 48 & 36 & 4, 10 & 2.9 (2.5--4.2) & 1 (2.7) & 323 & 2.8 \\
\bottomrule
\end{tabular}

\caption{Processed discharges, broken down by campaign, and within these by exposure setting. The ``In stats'' column shows only those discharges with at least 50 selected tracks. The ``pooled'' and ``own'' columns are the calibrations' scales presented in Section~\ref{sec:calibration}, with their 95\% bootstrap intervals. The columns $\tilde v_r$ and $\tilde\delta_r$ are the median over discharges of each discharge's per-discharge median velocity and width.}
\label{tab:campaigns}
\end{table}

The LCFS boundary contour, $q_{95}$, plasma current, magnetic axis, and vacuum toroidal field at the magnetic axis are taken from the EFIT reconstruction, sampled at 5~ms. The LCFS radius in the outboard midplane is taken as the maximum major radius that is within 5~cm of the magnetic axis height. This is the quantity that the camera limb tracks. The line averaged density is taken from the \texttt{esm} summary signal when available, otherwise it is taken from the CO$_2$ interferometer summary. The Greenwald fraction uses the Greenwald density $n_{\rm GW} = I_p/(\pi a^2)$, with plasma current $I_p$ in MA and minor radius $a$ from EFIT. Thomson scattering profiles are only available in the later campaigns and only reach the LCFS in a subset of the discharges within the flat top, \nTeMeasured{} of the \nShotsStats{} discharges used for statistical summaries.

\section{Method}
\label{sec:method}

\begin{figure}[t]
\centering
\includegraphics[width=\linewidth]{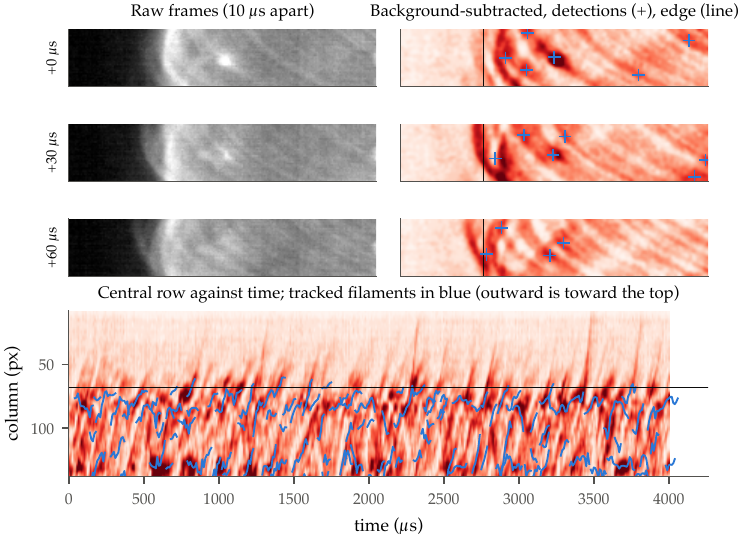}
\caption{Example of the pipeline for discharge 15235 in campaign M6 with an exposure of 7~$\mu$s. Top row: frames from the fast camera at 30~$\mu$s intervals, followed by the background-subtracted frames with detections marked with crosses and the detected edge column marked with a line. Bottom row: central row of the background-subtracted strip, plotted against time, with the detected filaments overlaid. As the SOL direction is up in this view, the filaments are inclined streaks moving in the SOL direction.}
\label{fig:overview}
\end{figure}

The stages of the pipeline for processing one discharge (or part of one) are shown in Figure~\ref{fig:overview}. All stages use the same settings for each shot; no tuning is undertaken on a case-by-case basis.

\subsection{Strip geometry and the plasma edge}
Each recording consists of $n$ frames of size $48 \times 256$ with 10-bit depth. Each column of the strip spans the plasma edge; the SOL and the core are at opposite ends of the strip depending on the mounting of the camera. For each frame we compute the profile of brightness $p(t,x)$ averaged over rows and smoothed in time with a window of 200 frames (2~ms) and spatial $\sigma = 2$ columns. The outer quarter of the columns is darker, and we identify this as the SOL. Thus we define a sign $s = \pm 1$. The velocity is then defined as $s\,\mathrm{d}x/\mathrm{d}t$ for outward motion.

We define two important positions $x_{\rm half}(t)$ and $x_{\rm peak}(t)$. The former is the first column from the SOL for which the brightness $p$ exceeds a threshold. The threshold is the midpoint between a ``dark level'' (average brightness of the ten outermost columns on the SOL side) and an ``inside level'' (median brightness of columns between 40\% and 75\% from the SOL side). We locate $x_{\rm half}$ by linear sub-pixel interpolation. $x_{\rm peak}$ is the brightest point within 40 columns from $x_{\rm half}$ on the inside. We perform a parabolic interpolation to refine the latter. We refer detections to $x_{\rm half}$ while processing and shift results to $x_{\rm peak}$ while aggregating. The peak is anchored to the equilibrium by the calibration presented in Section~\ref{sec:calibration}.

\subsection{Background subtraction and detection}
As filaments are transient, we estimate the steady background per pixel by the running minimum of the field over 21 frames ($\pm 10$ frames, or $\pm 100~\mu$s) \citep{kirk2016filaments}. The fluctuation is the difference between the frame and the background. We filter the fluctuation with a median filter of size $3 \times 3$ and a Gaussian filter with $\sigma = 1$ pixel. We estimate the noise $\sigma_n$ as 1.4826 times the median absolute deviation of all fluctuation pixels in a set of 3{,}000 frames. We find detection candidates as 8-connected components of pixels above $3\sigma_n$. We discard those with fewer than 12 pixels. We obtain the intensity-weighted centroid $(x_c, y_c)$ of each component as well as its peak fluctuation, its orientation and the major and minor axes of its second central moments. We report the chord FWHM as the width of the detection: the number of columns with fluctuation $\ge$ half the maximum in a 5-row average centred on $y_c$.

We measure width on a chord because of the slanted appearance of filaments. The column width of a component grows with its tilting. We report the signed SOL distance of a detection as $s\,(x_c - x_{\rm half})$.

\subsection{Tracking}
We link detections in successive frames using the Hungarian method \citep{kuhn1955hungarian}, minimizing the distance between centroids. We use a gate of 12 pixels per frame, corresponding to about 4.5~cm or 4.5~km/s at our calibrated scale. New tracks are started when there is no match, and they are ended when there is no match in a frame. We discard tracks with fewer than 5 detections (50~$\mu$s). The radial velocity of a track is calculated as the slope of a least squares linear regression of $s\,x_c$ against frame index with the corresponding standard error from the residuals. The width of a track is calculated as the median of the chord FWHM of its detections, and its SOL distance as the median of the signed distances.

We process recordings in blocks of 3{,}000 frames with an overlap of 60. We discard tracks started in the overlap with the preceding block from the results of the preceding block, but process and include them in the results of the following block.

\subsection{Optical flow as an independent velocity}
We verify our velocity estimate at each detection using Lucas--Kanade optical flow \citep{lucas1981iterative}. Over a $7 \times 7$ window centred on the centroid, we use the two Sobel spatial gradients and the difference between frames as a $2 \times 2$ least squares system for the displacement from frame $f$ to $f+1$. We solve the system whenever it has condition number lower than $10^4$. The flow velocity of a track is the mean of the horizontal components of flow over the track, with sign $s$. This estimate does not involve any assignment, but is responsive to the whole brightness profile in the neighbourhood of the centroid. Agreement between the two estimates is evidence that they are not merely methodological artefacts.

\subsection{ELM detection}
Edge-localized modes (ELMs) eject large filaments that would dominate the velocity distribution, so ELMs are detected from the midplane $D_\alpha$ monitor sampled at 50--250~kHz, depending on the campaign, with a spike defined to be a temporal peak whose residual after subtracting a 5~ms running median exceeds six times its robust standard deviation. The centre of mass in time of each spike's samples is recorded. Some campaigns provided a coarsely digitized version of the $D_\alpha$ monitor, with only tens of discrete values, making its robust standard deviation zero; in these cases the threshold is six digitizer steps, or half the inter-ELM level, whichever is larger.

The $D_\alpha$ channel is missing or saturated in some discharges, so we also apply the detector to the frame means of camera brightness, flagging a track if it starts within 1~ms of any spike. Discharges with more than five spikes in the $D_\alpha$ monitor in the camera time window are labelled as ELMy for the purposes of this paper, and tracks with ELM flags are removed from all statistical analyses. The parameter dependences of Section~\ref{sec:analysis} are all calculated from discharges that are not ELMy.

\subsection{Pixel scale from the equilibrium}
\label{sec:calibration}
The strip recordings in the archive are not spatially calibrated; this scale must be derived from the data. The limb peak position is where the line of sight is tangent to the LCFS, so the EFIT outboard radius of the LCFS, $R_{\rm LCFS}$, should track the sensor column of the peak position. The derivative of the median (over time) column versus $R_{\rm LCFS}$ across discharges gives the inverse scale. Within a discharge, however, $R_{\rm LCFS}$ changes only a few centimetres during the flat top, and EFIT has large error bars during ramp-up and termination. Also, the rising density changes the brightness profile shape, so regression within a shot is unstable (Appendix~\ref{app:calibration}). On the other hand, the plasma moves 10--20~cm between shots within a campaign, while the camera and its optics do not move within a campaign. This gives a long lever arm for a more stable regression.

For every discharge, we record $R_{\rm LCFS}$ at every EFIT time in the flat top. $R_{\rm LCFS}$ is the EFIT boundary contour's maximum major radius within 5~cm of the magnetic axis height, and $x_{\rm peak}$ is taken from the 2~ms smoothed profile. Any EFIT sample with $R_{\rm LCFS}$ more than 8~cm different from the flat-top median is discarded as unconverged. This gives one data point per discharge: the pair of median $x_{\rm peak}$ and $R_{\rm LCFS}$ for the discharge. To convert $x_{\rm peak}$ to sensor coordinates, we simply add the region-of-interest offset from the camera metadata. We define a camera configuration as (campaign, view string, region-of-interest offset), since any change in these three components could mean a different lens, pointing angle, or sensor region.

If a camera configuration has at least 8 discharges with a spread of at least 3~cm in $R_{\rm LCFS}$, we compute its own Theil--Sen slope \citep{sen1968estimates}; otherwise we use a pooled value from its campaign. The configuration slope is kept if it agrees in sign with $s$, has Spearman correlation $|\rho| \ge 0.5$, and its bootstrap 95\% interval excludes 0; otherwise a pooled slope is used instead. To get the pooled campaign slope, we regress column on $R_{\rm LCFS}$ after subtracting each configuration's mean $R_{\rm LCFS}$ and mean column to remove per-configuration intercepts. Bootstrap 95\% intervals are obtained from 2{,}000 bootstrap resamples of discharges within configurations \citep{efron1993bootstrap}. The scale is the inverse of the absolute value of the slope.

\subsection{Physical units and track selection}
Radial velocity is $v_r = s\,\dot x_c \times f_{\rm frame} \times \ell$, and radial width is $\delta_r = \mathrm{FWHM} \times \ell$, with $\ell$ the pixel scale; we measure distance from $x_{\rm peak}$ using the per-shot median offset $x_{\rm peak} - x_{\rm half}$. A track is included in statistics if its median position is within 10 pixels inboard and 40 pixels outboard of the limb peak (3--5~cm inboard and 11--21~cm outboard, depending on configuration scale), its velocity standard error is less than 0.5 pixels/frame, its width is finite, it starts inside the plasma flat top, and it is not ELM flagged. A discharge is included in shot-level statistics if it contains at least 50 tracks with $v_r > 0$, ensuring stable per-shot median calculations.

\subsection{Statistics}
Central values are medians, with 95\% bootstrap intervals from 2{,}000 resamples. Slopes are Theil--Sen estimates in $\log_{10}$ versus $\log_{10}$ space, with 95\% estimator intervals and the Spearman $\rho$ statistic. The velocity--size slope is calculated both for each discharge individually and pooled. For the latter, we normalize each track by dividing $v_r$ and $\delta_r$ by the discharge median values, so as not to introduce an artificial correlation in the pool because of differences between the discharges, like the camera pixel scale (e.g. different camera configurations) or plasma conditions. The exposure blur correction replaces $\delta_r$ by $\delta_{\rm corr} = (\delta_r^2 - (v_r t_{\rm exp})^2)^{1/2}$, then re-fits the slope, treating the smear as a Gaussian with width $v_r t_{\rm exp}$ added in quadrature.

Slopes are Theil--Sen estimates of medians of $v_r$ and $\delta_r$ for each discharge, versus plasma current, Greenwald fraction and injected beam power. The beam power has an offset of 0.05~MW to allow fitting of the Ohmic discharges in the log fit.

The two-region normalization requires $T_e$ and $n_e$ at the LCFS, $B$, $q_{95}$ and $R$. $T_e$ is taken from a Thomson scattering profile, the edge system if available, else the core system. The profile is interpolated at $R_{\rm LCFS}$, and $T_e$ is taken as the median value during the flat top. If no profile is available, $T_e$ is assumed to be \teDefault{}~eV and marked as such. The separatrix density $n_{e,\rm sep}$ is one-half of the line-averaged density. $B$ is the vacuum toroidal field at the magnetic axis, scaled by $R_{\rm axis}/R_{\rm LCFS}$. The connection length is $L_\parallel = \pi q_{95} R_{\rm axis}$. We take the model filament radius as half the measured FWHM, i.e.\ $a = \delta_r/2$. The collision frequency $\nu_{ei}$ is taken from the NRL formulary \citep{nrl2019formulary} with deuterium ions.

\section{Results}
\label{sec:results}

\subsection{Pixel scale}
\begin{figure}[t]
\centering
\includegraphics[width=\linewidth]{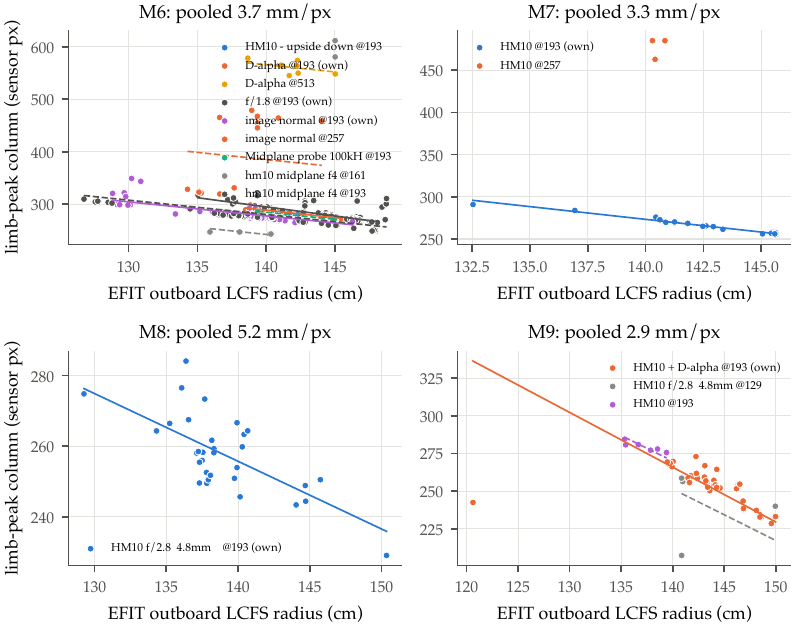}
\caption{Shot-to-shot pixel calibration. Each point is a discharge, plotted at its median position of the limb peak as a camera column, against the outboard LCFS radius from EFIT. Different colours represent different camera configurations, which differ in view string and region-of-interest offset. Solid lines are Theil--Sen fits per configuration, and dashed lines give the pooled campaign slope combined with each configuration's intercept. The scale is 1/slope.}
\label{fig:calibration}
\end{figure}

Of the \nConfigs{} camera configurations listed in Table~\ref{tab:configs}, \nConfigsOwn{} have enough discharges and spread in $R_{\rm LCFS}$ for individual fitting. These six configurations cover \nShotsOwnScale{} of the \nShotsProcessed{} discharges processed. For the other configurations, we obtained a pooled slope fit with all discharges of that campaign (Figure~\ref{fig:calibration}). In all of these fits, the limb peak column falls linearly with $R_{\rm LCFS}$, with Spearman correlations between \scaleOwnRhoWeak{} and \scaleOwnRhoStrong. For all slopes, the sign is consistent with the SOL direction seen on the images. The fitted scale ranges from \scaleOwnMin~mm\,pixel$^{-1}$ to \scaleOwnMax~mm\,pixel$^{-1}$. The tightest fit is M6 ``\scaleCfgCampSixView'' with \scaleCfgCampSixN{} discharges, $\rho = \scaleCfgCampSixRho$ and \scaleCfgCampSix~mm\,pixel$^{-1}$ (\scaleCfgCampSixLo--\scaleCfgCampSixHi). The M9 configuration is \scaleCfgCampNine~mm\,pixel$^{-1}$ (\scaleCfgCampNineLo--\scaleCfgCampNineHi).

\citet{farley2019identification} quote a resolution of 5~mm at the tangency plane for wide-angle images, reduced to $256 \times 160$ pixels from the camera's full $640 \times 448$ mode, i.e.\ about 2~mm per native pixel. This is the same order of magnitude as in M9, but 2~mm is outside M9's interval. The archive, however, does not store whether the camera was fitted with the same lens in M9.

We must note two limitations of our calibration. The 10--30\% intervals are large enough to hide bias in specific configurations, e.g.\ if the emission peak is shifted from the LCFS depending on density. The filaments also appear visually similar in pixel units for all configurations: across the \nCfgPx{} fitted configurations with at least 10 discharges each, the median track velocity is \vxPxCfgMin--\vxPxCfgMax{} pixels\,frame$^{-1}$ and width \fwhmPxCfgMin--\fwhmPxCfgMax{} pixels, yet the fitted scales range by a factor of \scaleOwnRatio.

In Table~\ref{tab:campaigns}, the differences between campaigns, which reflect their respective calibrations, are left uninterpreted. A scale error scales both $v_r$ and $\delta_r$ of a configuration by the same factor, so the velocity--size slopes of Section~\ref{sec:analysis} are unaffected, as $v_r$ and $\delta_r$ are normalized to the discharge median values before fitting. The diagonal offset of a discharge in the regime diagram would not affect the conclusions there.

\subsection{Velocity and width distributions}
\begin{figure}[t]
\centering
\includegraphics[width=\linewidth]{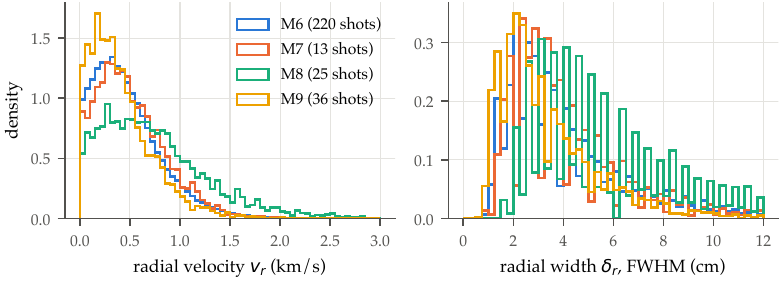}
\caption{Radial velocity (left) and radial FWHM (right) distributions for the selected track set. Data are pooled across discharges for each campaign.}
\label{fig:distributions}
\end{figure}

For these analyses we use tracks from \nShotsStats{} discharges with $\ge 50$ tracks, containing a total of \nTracksSel{} tracks. The pooling of radial velocities and widths across all these discharges yields the distributions in Figure~\ref{fig:distributions}. The median velocity of all discharges is \vMedAll~m/s (95\% CI \vMedAllLo--\vMedAllHi~m/s); the median width is \dMedAll~cm (95\% CI \dMedAllLo--\dMedAllHi~cm). Both distributions have long tails, with velocities out to 2--3~km/s and widths up to 10~cm with a peak at 2--3~cm. The median fraction of tracks moving outward (retained by our near-edge quality cuts) is \outwardMed\%; the inward-moving tracks are concentrated inboard of the limb peak, where streaks move with the plasma rotation. For simplicity, we thus include only the outward-moving tracks in the velocity--size analysis.

\subsection{Independent velocity estimates}
We can estimate the radial velocity without tracking, by correlating fluctuations between pairs of consecutive frames in a near-edge window. We find a median velocity of \xcorrVMed~m/s across the \nXcorr{} discharges with usable correlations, \xcorrRatioMed$\times$ the tracking median (95\% interval \xcorrRatioLo--\xcorrRatioHi). Discharge medians correlate with tracking at Spearman $\rho = \xcorrShotRho$. The sub-unity ratio is expected, since cross-correlation also weights the slow inboard portions of streaks and the stationary limb.

The Lucas--Kanade flow at detection points is another way to estimate velocities. It correlates with track velocities (median per-discharge correlation \lkAgreeMed), and Spearman $\rho = \lkShotRho$ across discharges at the medians. However, it is only \lkRatioMed{} times the track velocity, since the extended, inclined streaks impose the brightness-constancy constraint only normal to them, and any displacement $>1$ pixel/frame along with this constraint biases the flow magnitude low. Thus we use it for the sake of verifying the sign and ordering of velocities only.

\section{Analysis}
\label{sec:analysis}

\subsection{Velocity against size}
\begin{figure}[t]
\centering
\includegraphics[width=0.55\linewidth]{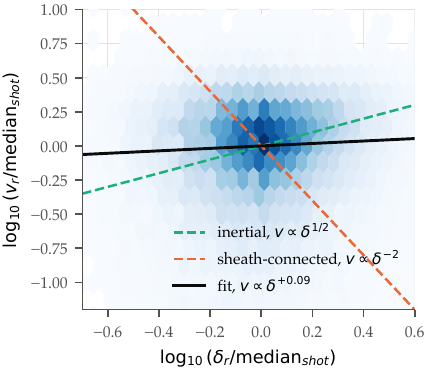}
\caption{Radial velocity against radial width for each selected track, normalized by the discharge median and plotted as a hexbin density. The solid line is the pooled Theil--Sen fit. The dashed lines show the limiting scaling for plasma connected through the sheath and for inertial flow, passing through the medians.}
\label{fig:vd}
\end{figure}

In Figure~\ref{fig:vd} we plot radial velocity against width for each track, normalized by the median for the discharge in which it was detected. We find a median per-discharge Theil--Sen slope of \vdSlopeMed{} (\vdSlopeMedLo{} to \vdSlopeMedHi), positive in \vdSlopePosShare\% of discharges. The pooled estimate is \vdSlopePooled{} (\vdSlopePooledLo{} to \vdSlopePooledHi) at Spearman $\rho = \vdRhoPooled$. The observed relation departs significantly from the limiting scaling $\propto \delta^{-2}$ for plasma connected through the sheath, and lies well below the inertial scaling $\propto \delta^{1/2}$.

\subsection{Exposure blur}
Filament motion during the exposure time blurs their image in the direction of motion, with greater velocity smearing larger distances. This introduces a spurious positive velocity--width correlation. We find the median smear distance $v_r t_{\rm exp} = \blurMedCm$~cm, only about 1/10th of the median width, so the effect is small but measurable. We thus use two tests to bound this potential bias. First, we observe in M6 a rise of slope with longer exposure time, from \vdSlopeExpSeven{} (\vdSlopeExpSevenLo{} to \vdSlopeExpSevenHi) at 7~$\mu$s ($n = \nExpSeven$), to \vdSlopeExpTen{} (\vdSlopeExpTenLo{} to \vdSlopeExpTenHi) at 10~$\mu$s ($n = \nExpTen$). Median width remains \dMedExpSeven~cm for both exposures, since the additional 0.1~cm smear over the 3~$\mu$s interval, at the median speed of 400~m/s, is negligible.

The 4~$\mu$s exposures lie in M7 and M8, so we cannot compare their widths with M6, and the slope is also low at \vdSlopeExpFour{} (\vdSlopeExpFourLo{} to \vdSlopeExpFourHi). Second, we find that correcting widths for their own smear in quadrature reduces the pooled slope to \vdSlopePooledBlur{} (\vdSlopePooledBlurLo{} to \vdSlopePooledBlurHi), which we take as our measurement. This shows that radial velocity is essentially independent of width over our measured range, with only a possibly slight positive correlation.

\begin{figure}[t]
\centering
\includegraphics[width=\linewidth]{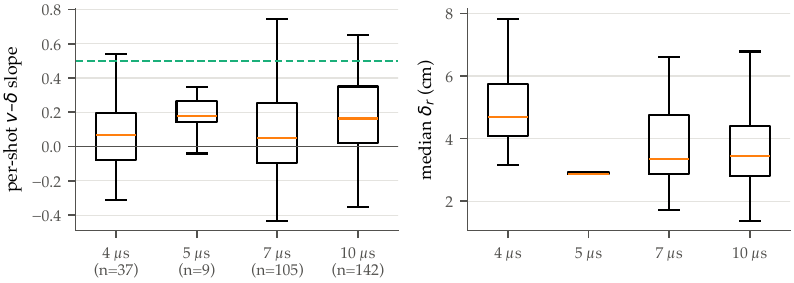}
\caption{Slopes of velocity vs.\ width per discharge for each exposure, with the inertial-scale relation as a reference (dashed) (left). Median width per discharge and exposure setting (right).}
\label{fig:exposure}
\end{figure}

\subsection{Two-region normalization}
\begin{figure}[t]
\centering
\includegraphics[width=0.55\linewidth]{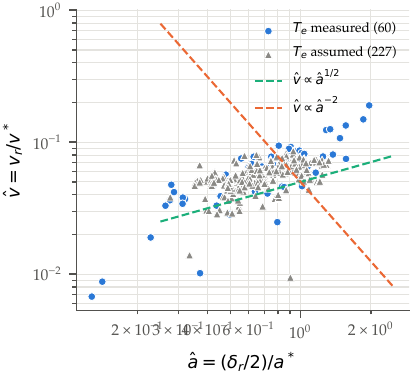}
\caption{Per-discharge median velocity and size in two-region units, with the inertial branch ($\hat v = \hat a^{1/2}$) and sheath-connected branch ($\hat v = \hat a^{-2}$) as reference (dashed). Discharges with measured electron temperature at the LCFS are marked with circles, while those with an assumed $T_e = \teDefault$~eV are marked with triangles.}
\label{fig:regimes}
\end{figure}

By applying the inputs for our discharges, as described in Section~\ref{sec:method}, we find the median scales to be $a^* = \aStarMed$~cm, $v^* = \vStarMed$~m/s, and $\Lambda = \LambdaMed$, placing our discharges in the low-collisionality region of the regime diagram. In Figure~\ref{fig:regimes}, we have plotted each discharge's median position on the $\hat a$--$\hat v$ diagram ($\hat a = (\delta_r/2)/a^*$, $\hat v = v_r/v^*$). We find $\hat a$ in the range \aHatMin{} to \aHatMax, with a median of \aHatMed. This range covers the peak of the model's velocity curve (near $\hat a = 1$), and the model's $\hat v(\hat a)$ remains within a factor of two of its peak across the middle half of $\hat a$'s range in the model's Eq.~(\ref{eq:regimes}). We thus expect a weak velocity--size relationship, as confirmed by our data.

However, the normalized velocity is only about \vHatMed, so our filaments move at just a few percent of the model's characteristic velocity. This discrepancy partly reflects our input assumptions, since $v^* \propto T_e^{7/10}$ and we take $T_e$ at the LCFS to be the assumed value for \nTeAssumed{} of our \nShotsStats{} discharges. The model also assumes an isolated, fully formed filament, and includes an order-unity prefactor. This prefactor is found to be below unity in simulations of realistic geometry \citep{walkden2015characterization, paruta2018blob}.

Furthermore, our measurements are made within 15~cm from the limb, where the filaments are still forming and accelerating and may not yet have reached the velocity described by the model. The gap is larger than is expected from the uncertainty in the prefactor alone, so the central disagreement with the model concerns the velocity's magnitude, not its size dependence.

\subsection{Dependence on plasma parameters}
\begin{figure}[t]
\centering
\includegraphics[width=\linewidth]{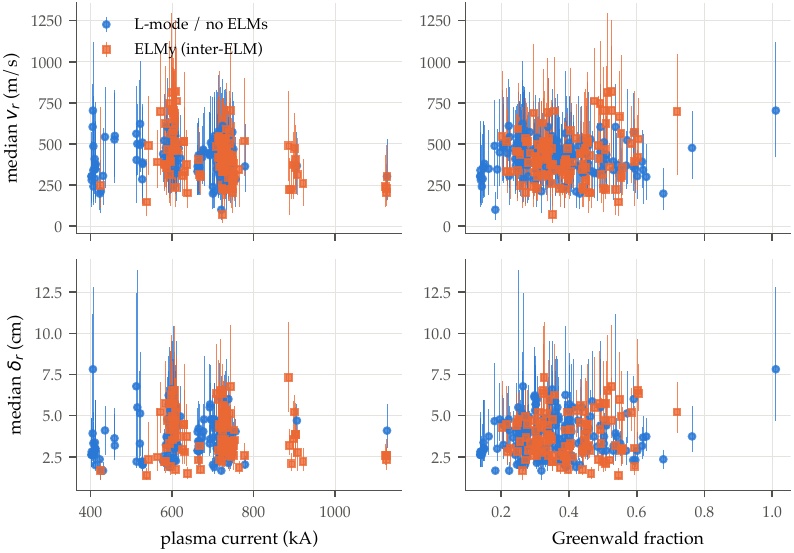}
\caption{Medians across each discharge of radial velocity (top) and width (bottom) versus plasma current (left) and Greenwald fraction (right). The bars are the interquartile range (IQR) across tracks; the circles are discharges without any ELMs in the camera window; the squares are ELMy discharges measuring inter-ELM velocities and sizes.}
\label{fig:params}
\end{figure}

In Figure~\ref{fig:params}, we show the per-discharge medians of velocity and width vs.\ plasma current and Greenwald fraction for our \nShotsL{} non-ELMy discharges. The discharges span a range of \ipMin{} to \ipMax~kA of plasma current and \fgwMin{} to \fgwMax{} of Greenwald fraction. We see no clear dependence on either current or Greenwald fraction. The log--log slopes of our median velocity distributions are \slVIp{} on current (95\% interval \slVIpLo{} to \slVIpHi) and \slVFgw{} on Greenwald fraction (\slVFgwLo{} to \slVFgwHi); those of the median width distributions are \slDIp{} on current (\slDIpLo{} to \slDIpHi) and \slDFgw{} on Greenwald fraction (\slDFgwLo{} to \slDFgwHi). The log--log slopes of median velocity and width on beam power are \slVNbi{} (\slVNbiLo{} to \slVNbiHi) and \slDNbi{} (\slDNbiLo{} to \slDNbiHi), respectively. The fits for the medians of our global sample combine multiple configurations with different calibration settings; a single, best-calibrated configuration, ``\bestCfgView,'' applies to a smaller subset of \nShotsBestCfgL{} non-ELMy discharges, at a single scale setting and plasma currents of \ipBestMin{} to \ipBestMax~kA. When we re-run the fits for this subset, we find velocity slopes of \slBVIp{} (\slBVIpLo{} to \slBVIpHi) on current and \slBVFgw{} (\slBVFgwLo{} to \slBVFgwHi) on Greenwald fraction, in agreement with our global result.

\citet{kirk2016filaments} report a current dependence of radial velocity falling with increasing current, but they study a dedicated current scan at constant density and power. In our sample, the current varies concurrently with density, heating, and research campaign. Moreover, most of our discharges are clustered around 700~kA. Still, a weak dependence might be masked by this confounding. The reported slopes describe our measurements, and are not intended to contradict the earlier result.

\section{Related Work}
\label{sec:related}

\citet{dudson2008experiments} studied L-mode discharges using MAST images and BOUT simulations, finding that the mode number of edge turbulence increases with density and decreases with $q_{95}$ and the filament width has the opposite trend. Their simulated filaments are born within about 2~cm of the edge, in the self-generated $E \times B$ shear layer. \citet{benayed2009interelm} looked at filaments in the inter-ELM interval, using both the camera and a reciprocating Langmuir probe, finding that filaments rotate near the LCFS and drift radially outward, that at higher density there are more and faster filaments, and used the radial velocities to test theoretical scaling relations for sheath-connected filaments and inertial filaments.

\citet{kirk2016filaments} scanned the plasma current at fixed density and heating, again in L-mode, and found that the radial velocity and (to a lesser extent) the radial size of filaments decrease with increasing plasma current, consistent with measured divertor profiles. \citet{farley2019identification} did a tomographic inversion of wide-angle images onto equilibrium field lines, recasting the problem of identifying filaments to one of field-aligned identification. All these studies relied on a camera geometry model built with Calcam \citep{silburn2018calcam}.

Our work differs in three ways. First, we don't use wide-angle frames; we use strip mode frames which have only 48 rows of pixels, but are taken at 100~kHz. Second, we derive the pixel scale from the equilibrium model rather than the geometry model. Third, we analyzed hundreds of discharges across four different campaigns; the other studies were limited to a few discharges.

The sheath-connected scaling is due to \citet{krasheninnikov2001scrape}. The inertial scaling is due to \citet{garcia2006fluctuations}. The two-region normalization model used in this paper was introduced by \citet{myra2006collisionality}; the model and its comparison with experiments are reviewed in \citet{dippolito2011convective}. The velocity--size relations were also measured on TORPEX, and the predicted transitions between regimes were observed \citep{theiler2009blob}. 3D simulations of filaments \citep{easy2014three, walkden2015characterization, paruta2018blob} reproduce these scaling relations in form, though realistic geometry and parallel dynamics affect the prefactors. \citet{zweben2007edge} review the imaging of edge turbulence in various devices, including how image intensity relates to density fluctuations.

We used standard detection and tracking methods: connected component labelling after robust thresholding, frame-to-frame assignment according to the Hungarian method \citep{kuhn1955hungarian}, velocity checks using Lucas--Kanade optical flow \citep{lucas1981iterative}, the Theil--Sen estimator \citep{sen1968estimates} for the slope, and the bootstrap \citep{efron1993bootstrap} for confidence intervals. All data came from the FAIR-MAST archive \citep{jackson2024fairmast}.

\section{Limitations}
\label{sec:discussion}

We derive the pixel scale from the limb peak tracking the radius of the outboard LCFS from the EFIT equilibrium with a fixed offset. This offset may drift with changes to the emission profile which depend on density and neutral pressure, and the EFIT boundary has uncertainty on the order of a centimetre. The bootstrap intervals of Table~\ref{tab:configs} include the scatter of these effects, though we did not account for a potential configuration-wide bias. The filaments appear to have very similar pixel sizes across configurations whose pixel scales vary by almost a factor of two (Section~\ref{sec:results}). Whether this is a coincidence or a sign that some of the spread in scale is not purely optical, we are unable to resolve at present with the archive; none of our conclusions depends on comparing physical units between configurations.

Every value of velocity or width has a systematic uncertainty of 10--30\% from the pixel scale of its configuration, and up to a factor of two uncertainty when comparing between configurations. This uncertainty could be removed by having a per-shot Calcam model, which is the obvious next step if the calibration images are made available. The failure of within-shot regression as a surrogate for Calcam is shown in Appendix~\ref{app:calibration}.

\paragraph{Geometry correction.}
The observed strip is one row band crossing the edge, whose column axis is radial only in the tangency plane. The measured width is the chord length of an inclined streak, which overestimates the perpendicular width by $1/\cos\theta$, where $\theta$ is the angle between the streak and the column axis. We do not correct this because the angle is hard to measure on a 48-row band, and roughly constant over the campaign. This shifts the $\hat a$ parameter but does not affect the velocity--size slope.

\paragraph{Emission blurring.}
The measured emission is integrated along the line of sight. A bright structure in the ``background'', beyond the tangency plane, is projected to a smaller radius, which blurs the SOL distance measurement. This does not affect the velocity because it is a time derivative.

\paragraph{What the camera sees.}
The $D_\alpha$ emission depends on the neutrals as well as plasma density. The neutrals usually fall off sharply inside the LCFS. This means filaments are visible mainly in the SOL. Filament brightness therefore changes as filaments move. Tracking the centroid is robust to the slowly varying amplitude, but optical flow is sensitive to amplitude changes. This is one reason for the low measured flow velocities.

\paragraph{Selection.}
The region of measurement, 10 pixels inboard to 40 pixels outboard, is 3--5~cm inward and 11--21~cm outward (depending on the configuration). This contains filaments at their birth, meaning the velocities in the far SOL may be different. The selection threshold of 50 tracks per discharge keeps the per-discharge medians stable. Tracks near ELMs are ignored. This discards some discharges and some of the largest events. The headline results refer to inter-ELM and L-mode filaments near the edge.

\paragraph{Model inputs.}
The LCFS temperature is assumed for most discharges, and the separatrix density $n_{e,\rm sep}$ is half the line average. Both enter the dimensionless parameters $a^*$, $v^*$, and $\Lambda$ as power laws with exponents less than unity. A factor of 2 uncertainty in either parameter means a discharge moves by less than a factor of 2 in the regime diagram. The conclusion that the sample covers $\hat a = 1$ is robust to these uncertainties. Note that the dimensionless velocity $\hat v$ itself is only known to within a factor of 2.

\section{Conclusion}
\label{sec:conclusion}
Public archives allow filament statistics to be measured on hundreds of discharges using an automated pipeline. MAST 100~kHz strip data was used to extract radial velocities and widths for \nTracksSel{} filaments in \nShotsStats{} discharges. A pixel scale is recovered based on the equilibrium to 10--30\% accuracy per camera configuration. Filament velocities have a median of \vMedAll~m/s, and a dependence on filament width after exposure blur correction with a log--log slope no larger than about 0.1. The filaments are located near the velocity maximum in two-region units, meaning that the velocity is nearly constant. However, the velocities are only a few percent of the velocity scale predicted by the model.

Three areas where a focused experiment or public release of a geometry model would improve upon the archives are the amplitude discrepancy, the campaign-level calibration, and the radial-only measurement geometry.

\section*{Acknowledgements}
The authors would like to thank the computer vision professors at Stanford, especially Juan Carlos Niebles, Adrien Gaidon, and Silvio Savarese for their courses that led to the ideas in this work.

\bibliography{refs}

\begin{thebibliography}{20}
\providecommand{\natexlab}[1]{#1}
\providecommand{\url}[1]{\texttt{#1}}
\expandafter\ifx\csname urlstyle\endcsname\relax
  \providecommand{\doi}[1]{doi: #1}\else
  \providecommand{\doi}{doi: \begingroup \urlstyle{rm}\Url}\fi

\bibitem[Ben~Ayed et~al.(2009)Ben~Ayed, Kirk, Dudson, Tallents, Vann, Wilson,
  and {the MAST Team}]{benayed2009interelm}
N.~Ben~Ayed, A.~Kirk, B.~Dudson, S.~Tallents, R.~G.~L. Vann, H.~R. Wilson, and
  {the MAST Team}.
\newblock Inter-{ELM} filaments and turbulent transport in the {Mega-Amp
  Spherical Tokamak}.
\newblock \emph{Plasma Physics and Controlled Fusion}, 51\penalty0
  (3):\penalty0 035016, 2009.
\newblock \doi{10.1088/0741-3335/51/3/035016}.

\bibitem[D'Ippolito et~al.(2011)D'Ippolito, Myra, and
  Zweben]{dippolito2011convective}
D.~A. D'Ippolito, J.~R. Myra, and S.~J. Zweben.
\newblock Convective transport by intermittent blob-filaments: Comparison of
  theory and experiment.
\newblock \emph{Physics of Plasmas}, 18\penalty0 (6):\penalty0 060501, 2011.
\newblock \doi{10.1063/1.3594609}.

\bibitem[Dudson et~al.(2008)Dudson, Ben~Ayed, Kirk, Wilson, Counsell, Xu,
  Umansky, Snyder, Lloyd, and {the MAST Team}]{dudson2008experiments}
B.~D. Dudson, N.~Ben~Ayed, A.~Kirk, H.~R. Wilson, G.~Counsell, X.~Xu,
  M.~Umansky, P.~B. Snyder, B.~Lloyd, and {the MAST Team}.
\newblock Experiments and simulation of edge turbulence and filaments in
  {MAST}.
\newblock \emph{Plasma Physics and Controlled Fusion}, 50\penalty0
  (12):\penalty0 124012, 2008.
\newblock \doi{10.1088/0741-3335/50/12/124012}.

\bibitem[Easy et~al.(2014)Easy, Militello, Omotani, Dudson, Havl{\'\i}{\v
  c}kov{\'a}, Tamain, Naulin, and Nielsen]{easy2014three}
L.~Easy, F.~Militello, J.~Omotani, B.~Dudson, E.~Havl{\'\i}{\v c}kov{\'a},
  P.~Tamain, V.~Naulin, and A.~H. Nielsen.
\newblock Three dimensional simulations of plasma filaments in the scrape off
  layer: A comparison with models of reduced dimensionality.
\newblock \emph{Physics of Plasmas}, 21\penalty0 (12):\penalty0 122515, 2014.
\newblock \doi{10.1063/1.4904207}.

\bibitem[Efron \& Tibshirani(1993)Efron and Tibshirani]{efron1993bootstrap}
Bradley Efron and Robert~J. Tibshirani.
\newblock \emph{An Introduction to the Bootstrap}.
\newblock Chapman and Hall/CRC, 1993.

\bibitem[Farley et~al.(2019)Farley, Walkden, Militello, Sanna, Young, Silburn,
  Harrison, Kogan, Lupelli, Henderson, Kirk, and
  Bradley]{farley2019identification}
T.~Farley, N.~R. Walkden, F.~Militello, M.~Sanna, J.~Young, S.~S. Silburn,
  J.~Harrison, L.~Kogan, I.~Lupelli, S.~S. Henderson, A.~Kirk, and J.~W.
  Bradley.
\newblock Filament identification in wide-angle high speed imaging of the {Mega
  Amp Spherical Tokamak}.
\newblock \emph{Review of Scientific Instruments}, 90\penalty0 (9):\penalty0
  093502, 2019.
\newblock \doi{10.1063/1.5109470}.

\bibitem[Garcia et~al.(2006)Garcia, Bian, and
  Fundamenski]{garcia2006fluctuations}
O.~E. Garcia, N.~H. Bian, and W.~Fundamenski.
\newblock Radial interchange motions of plasma filaments.
\newblock \emph{Physics of Plasmas}, 13\penalty0 (8):\penalty0 082309, 2006.
\newblock \doi{10.1063/1.2336422}.

\bibitem[Huba(2019)]{nrl2019formulary}
J.~D. Huba.
\newblock {NRL} plasma formulary.
\newblock Naval Research Laboratory, Washington, DC, 2019.

\bibitem[Jackson et~al.(2024)Jackson, Khan, Cummings, Hodson, de~Witt, Pamela,
  Akers, Thiyagalingam, and {the MAST Team}]{jackson2024fairmast}
Samuel Jackson, Saiful Khan, Nathan Cummings, James Hodson, Shaun de~Witt,
  Stanislas Pamela, Rob Akers, Jeyan Thiyagalingam, and {the MAST Team}.
\newblock {FAIR-MAST}: A fusion device data management system.
\newblock \emph{SoftwareX}, 27:\penalty0 101869, 2024.
\newblock \doi{10.1016/j.softx.2024.101869}.

\bibitem[Kirk et~al.(2016)Kirk, Thornton, Harrison, Militello, Walkden, {the
  MAST Team}, and {the EUROfusion MST1 Team}]{kirk2016filaments}
A.~Kirk, A.~J. Thornton, J.~R. Harrison, F.~Militello, N.~R. Walkden, {the MAST
  Team}, and {the EUROfusion MST1 Team}.
\newblock L-mode filament characteristics on {MAST} as a function of plasma
  current measured using visible imaging.
\newblock \emph{Plasma Physics and Controlled Fusion}, 58\penalty0
  (8):\penalty0 085008, 2016.
\newblock \doi{10.1088/0741-3335/58/8/085008}.

\bibitem[Krasheninnikov(2001)]{krasheninnikov2001scrape}
S.~I. Krasheninnikov.
\newblock On scrape off layer plasma transport.
\newblock \emph{Physics Letters A}, 283\penalty0 (5--6):\penalty0 368--370,
  2001.
\newblock \doi{10.1016/S0375-9601(01)00252-3}.

\bibitem[Kuhn(1955)]{kuhn1955hungarian}
H.~W. Kuhn.
\newblock The {H}ungarian method for the assignment problem.
\newblock \emph{Naval Research Logistics Quarterly}, 2\penalty0
  (1--2):\penalty0 83--97, 1955.

\bibitem[Lucas \& Kanade(1981)Lucas and Kanade]{lucas1981iterative}
Bruce~D. Lucas and Takeo Kanade.
\newblock An iterative image registration technique with an application to
  stereo vision.
\newblock In \emph{Proceedings of the 7th International Joint Conference on
  Artificial Intelligence (IJCAI)}, pp.\  674--679, 1981.

\bibitem[Myra et~al.(2006)Myra, Russell, and
  D'Ippolito]{myra2006collisionality}
J.~R. Myra, D.~A. Russell, and D.~A. D'Ippolito.
\newblock Collisionality and magnetic geometry effects on tokamak edge
  turbulent transport. {I}. a two-region model with application to blobs.
\newblock \emph{Physics of Plasmas}, 13\penalty0 (11):\penalty0 112502, 2006.
\newblock \doi{10.1063/1.2364858}.

\bibitem[Paruta et~al.(2018)Paruta, Beadle, Ricci, and Theiler]{paruta2018blob}
Paola Paruta, C.~Beadle, P.~Ricci, and C.~Theiler.
\newblock Blob velocity scaling in diverted tokamaks: a comparison between
  theory and simulation.
\newblock Technical Report WP14ER-PR(18) 21313, EUROfusion, 2018.
\newblock Preprint.

\bibitem[Sen(1968)]{sen1968estimates}
Pranab~Kumar Sen.
\newblock Estimates of the regression coefficient based on {K}endall's tau.
\newblock \emph{Journal of the American Statistical Association}, 63\penalty0
  (324):\penalty0 1379--1389, 1968.

\bibitem[Silburn et~al.(2018)Silburn, Harrison, Smithies, Wynn, Farley, and
  Cavalier]{silburn2018calcam}
S.~Silburn, J.~Harrison, M.~Smithies, A.~Wynn, T.~Farley, and J.~Cavalier.
\newblock Calcam (version 1.9.4).
\newblock \url{https://github.com/euratom-software/calcam}, 2018.

\bibitem[Theiler et~al.(2009)Theiler, Furno, Ricci, Fasoli, Labit, M{\"u}ller,
  and Plyushchev]{theiler2009blob}
C.~Theiler, I.~Furno, P.~Ricci, A.~Fasoli, B.~Labit, S.~H. M{\"u}ller, and
  G.~Plyushchev.
\newblock Cross-field motion of plasma blobs in an open magnetic field line
  configuration.
\newblock \emph{Physical Review Letters}, 103\penalty0 (6):\penalty0 065001,
  2009.
\newblock \doi{10.1103/PhysRevLett.103.065001}.

\bibitem[Walkden et~al.(2015)Walkden, Dudson, Easy, Fishpool, and
  Omotani]{walkden2015characterization}
N.~R. Walkden, B.~D. Dudson, L.~Easy, G.~Fishpool, and J.~T. Omotani.
\newblock Numerical investigation of isolated filament motion in a realistic
  tokamak geometry.
\newblock \emph{Nuclear Fusion}, 55\penalty0 (11):\penalty0 113022, 2015.
\newblock \doi{10.1088/0029-5515/55/11/113022}.

\bibitem[Zweben et~al.(2007)Zweben, Boedo, Grulke, Hidalgo, LaBombard, Maqueda,
  Scarin, and Terry]{zweben2007edge}
S.~J. Zweben, J.~A. Boedo, O.~Grulke, C.~Hidalgo, B.~LaBombard, R.~J. Maqueda,
  P.~Scarin, and J.~L. Terry.
\newblock Edge turbulence measurements in toroidal fusion devices.
\newblock \emph{Plasma Physics and Controlled Fusion}, 49\penalty0
  (7):\penalty0 S1--S23, 2007.
\newblock \doi{10.1088/0741-3335/49/7/S01}.

\end{thebibliography}
\bibliographystyle{colm2026_conference}

\clearpage
\appendix
\section{Camera configurations}
\label{app:views}
Table~\ref{tab:configs} lists the camera configurations from the processed set, giving the view string in the camera metadata and the sensor column offset of the region of interest. It also lists the number of discharges, their spread in $R_{\rm LCFS}$, the Spearman correlation of limb peak column and $R_{\rm LCFS}$, and gives the scale at which the configuration can be fitted. It additionally specifies whether the configuration is fitted at its own scale or whether it uses the pooled campaign scale (see Section~\ref{sec:calibration}).

\begin{table}[h]
\centering
\scriptsize
\setlength{\tabcolsep}{4pt}\begin{tabular}{llrrrlll}
\toprule
Camp. & View string & ROI left & Shots & $\Delta R_{\rm LCFS}$ (cm) & $\rho$ & Scale (mm/px) & Used \\
\midrule
M6 & HM10, normal, f/1.8 & 193 & 70 & 13 & -0.94 & 2.9 (2.7--3.0) & own \\
M6 & HM10, normal & 193 & 61 & 17 & -0.89 & 3.7 (3.1--4.1) & own \\
M6 & hm10 midplane f4 & 193 & 59 & 22 & -0.47 & 3.9 (3.6--5.1) & pooled \\
M6 & HM10, normal, D$\alpha$ 10 nm & 193 & 30 & 7 & -0.88 & 3.5 (3.2--3.9) & own \\
M6 & HM10, normal & 257 & 14 & 10 & +0.34 & 0.9 (0.4--1.9) & pooled \\
M6 & HM10, normal, D$\alpha$ 10 nm & 513 & 7 & 6 & -- & -- & pooled \\
M6 & hm10 midplane f4 & 161 & 6 & 4 & -- & -- & pooled \\
M6 & HM10 - upside down & 193 & 3 & 1 & -- & -- & pooled \\
M6 & probe 100kHz & 193 & 3 & 6 & -- & -- & pooled \\
M6 & HM10, normal, D$\alpha$ 10 nm & 545 & 2 & 0 & -- & -- & pooled \\
M6 & HM10, normal, D$\alpha$ 10 nm & 225 & 1 & 0 & -- & -- & pooled \\
M6 & HM10, normal & 225 & 1 & 0 & -- & -- & pooled \\
M6 & probe view & 193 & 1 & 0 & -- & -- & pooled \\
M7 & HM10 & 193 & 22 & 13 & -0.88 & 3.3 (3.0--3.8) & own \\
M7 & HM10 & 257 & 3 & 1 & -- & -- & pooled \\
M8 & HM10 f/2.8 4.8mm nd 0.9 & 193 & 32 & 21 & -0.58 & 5.2 (3.6--11.5) & own \\
M9 & HM10 + D$\alpha$ filter & 193 & 38 & 29 & -0.82 & 2.7 (2.5--3.4) & own \\
M9 & HM10 & 193 & 6 & 4 & -- & -- & pooled \\
M9 & HM10 f/2.8 4.8mm & 129 & 4 & 9 & -- & -- & pooled \\
M9 & HM10 + D$\alpha$ filter & 129 & 1 & 0 & -- & -- & pooled \\
M9 & HM10 + D$\alpha$ filter & 513 & 1 & 0 & -- & -- & pooled \\
M9 & HM10 + D$\alpha$ filter & 577 & 1 & 0 & -- & -- & pooled \\
M9 & HM10 & 129 & 1 & 0 & -- & -- & pooled \\
\bottomrule
\end{tabular}

\caption{Camera configurations and their calibration. We use the configuration's own scale if it has at least 8 discharges, a spread of at least 3~cm in $R_{\rm LCFS}$, a correlation (of the correct sign) with $|\rho|$ of at least 0.5, and a bootstrap interval that excludes zero.}
\label{tab:configs}
\end{table}

\section{Within-shot calibration}
\label{app:calibration}
The within-shot regression exploits the few centimetres of motion of the plasma during a flat top. For the \nWithinShot{} discharges in our pool in which the LCFS moves by at least 2~cm, we find that the median scale is \withinScaleMed~mm/pixel, with interquartile range \withinScaleQLo--\withinScaleQHi~mm/pixel, with extreme values of $< 1$~mm and $> 10^{\withinScaleMaxLog}$~mm. This large spread may be due to three factors. First, the EFIT boundary measurements are least reliable at the beginning and end of a flat top, at which we get most movement; second, as density increases throughout the flat top, the brightness profile is deformed, causing the peak at a fixed $R_{\rm LCFS}$ to move; finally, the lever arm is small, so a small error of 1~cm in $R_{\rm LCFS}$ causes a large error of tens of percent in our slope. The between-shot regression of Section~\ref{sec:calibration} is much more stable, as we consider the variation of position over a campaign, which is typically 10--20~cm, rather than the few centimetres of intra-discharge motion.

\section{ELM detection}
\label{app:elm}
\begin{figure}[h]
\centering
\includegraphics[width=\linewidth]{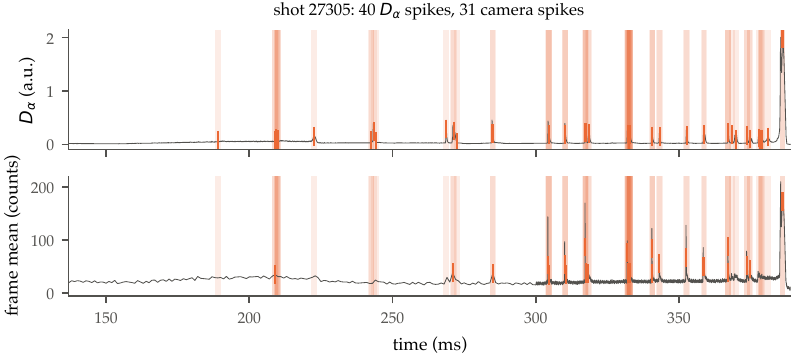}
\caption{ELM detection on an ELMy discharge. Upper panel: midplane $D_\alpha$ signal and its detected spikes. Lower panel: mean brightness of the camera frame and its detected spikes. The $\pm 1$~ms exclusion regions for both channels are shown in shading.}
\label{fig:elm}
\end{figure}

Figure~\ref{fig:elm} shows an ELMy discharge's $D_\alpha$ trace alongside the mean brightness of the camera frame. We also show the spikes detected by the camera, which is able to detect the same events as $D_\alpha$ when the latter are available, and which is our sole detector when the $D_\alpha$ channel is not available.

\end{document}